\documentstyle[preprint,prc,aps,epsfig]{revtex}
\begin{document}
\tighten
\draft
\title{ Rotating compact stars with exotic matter}
\author{Sarmistha Banik$^{\rm (a)}$, Matthias Hanauske$^{\rm (b)}$, Debades 
Bandyopadhyay$^{\rm (a)}$ and Walter Greiner$^{\rm (c)}$}
\address{$^{\rm (a)}$Saha Institute of Nuclear Physics, 1/AF Bidhannagar, 
Calcutta 700 064, India}
\address{$^{\rm (b)}$Institut f\"ur Theoretische Physik, J. W. Goethe 
Universit\"at, D-60054 Frankfurt am Main, Germany}
\address{$^{\rm (c)}$Frankfurt Institute for Advanced Studies (FIAS), 
J. W. Goethe Universit\"at, D-60054 Frankfurt am Main, Germany}

\maketitle

\begin{abstract}
We have constructed models of uniformly rotating compact stars including
hyperons, Bose-Einstein condensates of antikaons and quarks. First order
phase transitions from hadronic to antikaon condensed matter and then to quark 
matter are considered here. For the equation of state undergoing phase 
transitions to antikaon condensates, the third family of compact stars are
found to exist in the fixed angular velocity sequences. However, the third 
family solution disappears when the compact stars rotate very fast. For this
equation of state, the fixed baryon number supramassive sequence shows a second
stable part after the unstable region but no back bending phenomenon. On the  
other hand, we observe that the rotation gives rise to
a second maximum beyond the neutron star maximum for the equation of 
state involving phase transitions to both antikaon condensed and quark matter. 
In this case, the back bending phenomenon has been observed in the supramassive
sequence as a consequence of the first order phase transition from $K^-$ 
condensed to quark matter. And the back bending segment 
contains stable configurations of neutron stars.  

\end{abstract}
\section{Introduction}
The study of dense and cold matter in neutron star interior has gained momentum
after the advances in observational astronomy. One important goal of various
observations is to measure masses and radii of compact stars 
\cite{Pon,Dra,Wal,Cot}. This might, in turn,
constrain the composition and equation of state (EoS) of dense matter in 
compact stars. The composition and structure of a neutron star depend on the
nature of strong interaction. In neutron star cores, the matter density could
exceed by a few times normal nuclear matter density. Since the baryon and 
lepton chemical potentials increase rapidly with density in the core, different
exotic forms of matter with a large strangeness fraction such as hyperon 
matter \cite{Gle97}, Bose-Einstein condensates of strange mesons 
\cite{Kap,Pra97,Sch96,Gle99,Pal,Bani01,Bani02,Bani03,Bani04} and quark matter 
\cite{Gle97,Pra97,Far,Schr} may appear there. 

Neutron star matter including exotic phases was studied extensively using 
relativistic field theoretical models \cite{Wale} and chiral models 
\cite{Han00}. These studies have revealed many important aspects of the EoS of
dense matter and static neutron star properties. It was noted that 
the appearance of exotic forms of matter in the high density regime resulted in 
kinks in the EoS \cite{Bani02,Han02,Han03}. As a result, there was a 
discontinuity in the speed of
sound. It has a great implication. There is a growing interest to understand
whether a stable sequence of compact stars could exist in nature beyond a 
white dwarf and neutron star branch. Gerlach \cite{Ger} first showed that it 
could be a possibility if there was a jump in an EoS or discontinuity in the 
speed of sound. Later various groups constructed equations of state undergoing
first order phase transitions from hadronic to quark matter \cite{Han03,Ket}, 
hadronic to
antikaon condensed matter \cite{Bani02} and hadronic to hyperonic matter 
\cite{Han02} which gave rise to a stable branch of non rotating compact stars
called the third family beyond the neutron star branch. It was noted that 
nonidentical stars of same mass but different radii and compositions could 
exist because of partial overlapping mass regions of two branches. These pairs 
are called neutron star twins.        

On the other hand, the properties of rapidly rotating neutron stars may also 
constrain the EoS of dense matter. In particular, the maximum angular velocity 
of rotating neutron stars is quite sensitive to the EoS \cite{Noz,Bur}. An EoS 
predicting smaller Keplerian frequencies than the observed frequencies is not 
allowed. Also, it was earlier predicted that the back
bending phenomenon in a fixed baryon number evolutionary sequence could shed
light on the existence of quark matter in the neutron star interior
\cite{Web,Spy}. 

In this paper, we are interested to investigate the impact of exotic forms of 
matter on various properties of rotating compact stars. 
In this respect, we exploit equations of state including 
hyperons, Bose-Einstein condensates of strange mesons and quarks and derived
them from the relativistic field theoretical model and chiral model. 
The paper is organised in the following
way. In Section II, we discuss the space-time geometry of stationary and
axisymmetric rotating stars and their stability briefly. We describe
equations of state adopted in this calculation in Section III. In Section IV,
we discuss results of our calculation. Section V is devoted to summary and
conclusions.
\section{Uniformly Rotating Neutron Stars}
Models of rotating neutron stars were developed by various groups 
\cite{Wil,Bona,But,Fri,Fri2,KEH,KEH2,Cook0,Cook,Sal}. Here
we consider stationary and axisymmetric equilibrium configurations of rotating
neutron stars. In this case, the metric has the form \cite{Cook}
\begin{equation}
ds^2 = - e^{\gamma + \rho} dt^2 + e^{2\alpha} (dr^2 + r^2 d{\theta}^2)
       + e^{\gamma - \rho} r^2 sin^2{\theta}(d\phi - \omega dt)^2 ~,
\end{equation}
where metric potentials $\gamma$, $\rho$ and $\alpha$ and $\omega$ depend on
radial coordinate $r$ and polar angle $\theta$. The compact star matter is
considered to be a perfect fluid and the energy density and pressure are
determined from the stress-energy tensor. Here we construct models of
uniformly rotating neutron star using a code based on Komatsu-Eriguchi-Hachisu
method \cite{KEH,KEH2} and written by Steirgioulas \cite{Ster95} for three 
equations
of state. In this paper, we study both the static and mass shedding or 
Keplerian 
limit which is attained by a stable neutron star rotating with a maximum
frequency called the Kepler frequency $\Omega_K$ before the mass loss at the
equator occurs \cite{Cook}. Also, we focus our attention on rotating neutron
stars along evolutionary sequences where the total baryon number of a star
is conserved. There are two kinds of evolutionary sequences - normal and 
supramassive. Normal evolutionary sequence has one end connecting the static
limit whereas the supramassive evolutionary sequence does not contain any 
static solution.
Along an evolutionary sequence, the stability condition is given by 
\cite{Cook0,Cook}
\begin{equation}
{\frac {dJ} {d\epsilon_c}|}_{N_b = const.} < 0~,
\end{equation} 
where $J$ is the angular momentum, $\epsilon_c$ is the central energy density
and $N_b$ is the total baryon number.

\section{Equations of state}
In this paper, we use three equations of state constructed by us. 
One of these EoS is constructed from a 
QCD-motivated hadronic chiral SU(3)$_L \times$SU(3)$_R$ model \cite{Han00}. 
In this model, the Lagrangian is constructed with 
respect to the nonlinear realization of chiral SU(3)$_L \times$SU(3)$_R$ 
symmetry; masses of heavy baryons and mesons are generated by spontaneous 
symmetry breaking and masses of the pseudoscalar bosons are the results of
explicit symmetry breaking. A dilaton field is introduced to describe the 
gluon condensate. This EoS does not involve any phase transition and describes 
matter involving only neutrons, protons, hyperons and leptons. 
It is denoted as CHM case. This EoS was already used to study neutron star 
properties \cite{Han00,Shovy}. 

Next, we discuss the EoS undergoing a first order phase 
transition from hadronic to $K^-$ condensed matter and then a second order 
$\bar K^0$ condensation 
\cite{Bani02}. This is hereafter denoted as HK case. In the last case, the EoS
involves first order phase transitions from nuclear to $K^-$ condensed matter
and then from $K^-$ condensed matter to quark matter. It is referred to as HKQ 
case. In both cases, we adopt a relativistic field theoretical model to 
describe the pure hadronic and antikaon condensed matter and their mixed phase
\cite{Gle99,Bani02}. The constituents of the hadronic phase for HK case are the 
members of the baryon octet and leptons. On the other hand, the pure $\bar K$ 
condensed phase 
consists of members of the baryon octet, $K^-$, $\bar K^0$ mesons and leptons
for HK case. Similarly, the compositions of the hadronic and antikaon condensed 
phases for HKQ case are $n$, $p$, leptons and  
$n$, $p$, $K^-$ mesons and leptons respectively. 
In the condensed phase, baryons are embedded in the condensates. The 
baryon-baryon and baryon-(anti)kaon interactions are mediated by the scalar and
vector mesons. The model is also extended to include hyperon-hyperon
interaction \cite{Sch96}. On the other hand, the pure quark matter is composed 
of $u$, $d$, $s$
quarks and electrons. The quark phase is described by the MIT bag model. Each
pure phase maintains local charge neutrality and $\beta$-equilibrium 
conditions. The mixed phase of any two pure phases is described by the Gibbs
phase rules along with global charge neutrality and total baryon number
conservation. The meson-baryon and meson-kaon coupling constants used    
in our calculations for HK and HKQ cases are recorded in Ref.\cite{Bani02}.
An antikaon optical potential depth of $U_{\bar K} (n_0)$ = - 160 MeV at normal 
nuclear matter density is used in the calculation of EoS for HK and HKQ cases. 
Also, we use a bag constant $B^{1/4} =$ 200 MeV and strange quark mass $m_s =$
150 MeV for the quark matter. 

The structure of static compact stars was already investigated for
HK case using Tolman-Oppenheimer-Volkoff (TOV) equations \cite{Bani02}. For HK
case, it was noted that the threshold of $K^-$
condensation was very sensitive to the antikaon optical potential depth 
($U_{\bar K}(n_0)$) at normal nuclear matter density. We performed this 
calculation for a set of antikaon optical potential depths from -100 to -180 
MeV. For 
$|U_{\bar K}(n_0)| <$ 160 MeV, the early appearance of negatively charged 
hyperons
did not allow $K^-$ condensate to appear in the system. However, $K^-$ 
condensate appeared before the formation of negatively charged $\Xi$ hyperons
for $|U_{\bar K}(n_0)| \geq $ 160 MeV. So we performed the static structure 
calculation for HK case using $U_{\bar K}(n_0)$ = - 160 and -180 MeV. In this 
calculation, we found    
a new family of compact stars called the third family beyond the neutron star
branch \cite{Bani02}. After the maximum mass star in the neutron star branch, 
there
is an unstable region followed by a stable third family branch of superdense 
stars. However, we did not find this third family branch when we performed 
calculation with $U_{\bar K}(n_0)$ = - 180 MeV. So the third family branch
of compact stars appears in a certain range of antikaon optical potential 
depth. 
\section{Results and Discussion}
Now we present our results for CHM, HK and HKQ cases. In Fig. 1, equations
of state (pressure versus energy density) are displayed for those cases. 
For HK and HKQ cases, the mixed phase of hadronic and $K^-$ condensed
matter begins at energy density $\epsilon =$ 6.1132$\times 10^{14}$ g/cm$^3$ 
and 
ends at $\epsilon =$ 1.1651$\times 10^{15}$ g/cm$^3$. In HK case, $\Lambda$ 
hyperons appear in the mixed phase and $\bar K^0$ condensation sets in just 
after the mixed phase is over \cite{Bani02}. Also, heavier hyperons are 
populated 
at higher baryon density for HK case \cite{Bani02}. On the other hand, the first
order phase transition from $K^-$ condensed matter to quark matter for HKQ case
starts at $\epsilon =$ 1.5884$\times 10^{15}$ g/cm$^3$ and terminates at 
$\epsilon =$ 2.7688$\times 10^{15}$ g/cm$^3$. The overall EoS for HK case is 
softer than that of HKQ case whereas that of CHM model is stiffer than HK
case but softer compared with HKQ case.

The gravitational mass for static neutron star sequence and the sequence of 
neutron stars rotating at their respective Kepler frequencies are plotted with 
central energy density in Fig. 2 for CHM, HK and HKQ cases. 
The bottom three curves represent the 
static limit sequences calculated using TOV equations and the top three curves 
correspond to the mass shedding limit sequences for the above mentioned EoS. 
We have already mentioned that the EoS for HK case gives rise to the third
family of superdense stars beyond the neutron star branch. 
The maximum mass stars on the neutron star branch and the third family branch 
are indicated by solid and open circles respectively.
The maximum gravitational masses, equatorial radii and the 
corresponding central energy densities
for the static and mass shedding limit are given by Table 1. Also, the Kepler
periods for maximum masses in mass shedding limit sequences are recorded in 
the
Table. For HK case, the second row gives the relevant quantities for the third
family branch in the static sequence. The compact star in the third family has 
a smaller radius than the neutron star as it is evident from the Table. With the
help of the central energy densities corresponding to maximum mass static and 
rotating stars for HK and HKQ case from Table 1 and Fig. 1, we find that the 
corresponding central 
pressures for HK case are smaller than that of HKQ case. Therefore, the overall 
EoS from the center to the surface of a maximum mass star is softer in HK case 
compared with that of HKQ case. This leads to smaller maximum masses in the 
static and mass shedding limit sequence for HK case than that of HKQ case. 
 
The gravitational mass as a function of equatorial radius is shown in Fig. 3a,
Fig. 3b and Fig. 3c for CHM, HK and HKQ cases respectively. 
Also, we have plotted fixed angular velocity sequences in those
figures. Each fixed angular velocity curve is denoted by a value of $\Omega$. 
In Fig. 3a, fixed baryon number curve I with $N_b=1.71 \times 10^{57}$ and 
curve II with $N_b = 2.18 \times 10^{57}$ 
stand for normal sequences and curve III with $N_b=2.51 \times 10^{57}$ is the 
supramassive sequence. In Fig. 3b and 3c, fixed baryon number curves I and II 
with $N_b = 1.94 \times 10^{57} (2.15 \times 10^{57})$ 
and $2.06 \times 10^{57} (2.34 \times 10^{57})$ 
are normal sequences and curves III and IV with
$N_b = 2.09 \times 10^{57} (2.45 \times 10^{57})$ and 
$2.50 \times 10^{57} (2.57 \times 10^{57})$ 
are supramassive sequences for HK (HKQ) case respectively. 
In each figure, the solid circle exhibits the location of the maximum 
mass star in the neutron star branch. We note
that there are some interesting structures in the fixed angular velocity curves
for HK and HKQ cases which are undergoing first order phase transitions from 
hadronic to antikaon condensed and then to quark matter.

For HK case, the static sequence represented by the bold line at the
bottom in Fig. 3b has some interesting structure. After the maximum neutron 
star mass, there is an unstable region followed by a stable third family branch 
of compact stars \cite{Bani02}. The stellar mass peak in the third family 
branch is denoted by an open circle. So far some EoS 
undergoing first order phase transitions from hadronic to hyperon matter 
\cite{Han02}, antikaon condensed matter \cite{Bani02} and quark matter 
\cite{Schr,Han03,Ket} gave rise to the stable third family branch beyond the 
neutron star branch in the static case. Generally, it was noted that if the
neutron star maximum occurred in the mixed phase or just after that, there 
would be a third family branch because of the jump in the EoS \cite{Ger}. 

Now the question is whether the third family
branch could still exist under the influence of rotation. In Fig. 3b, we find 
that the third family branch exists for the fixed angular velocity sequence 
with $\Omega = 2158$ s$^{-1}$. But there is no third family solution for fixed 
angular velocity curves with $\Omega = 3000$ s$^{-1}$, $\Omega = 4900$ s$^{-1}$
and beyond. The disappearance of the third family branch may be understood in 
the following way. It was argued that the third family solution was very 
sensitive to the behaviour of EoS at higher density \cite{Ket}. 
The rotating stars have smaller central 
energy densities than the non rotating stars with the same gravitational mass.
This is attributed to the effect of the centrifugal force. It is also evident
from the higher angular velocity and mass shedding limit curves that the 
positions of maximum star masses move towards smaller central energy densities.
Consequently, the third family branch disappears because the high density part
of the EoS which is responsible for the third family solution, can not be probed
in fast rotating compact stars. 

On the other hand, there is no third family branch in the static limit
sequence for HKQ case in Fig. 3c. As angular velocity increases, the curves 
become more and more flattened due to the softening in the EoS by the antikaon
condensate and quarks. This leads to a local minimum as it is evident from  
fixed angular velocity sequences 
with $\Omega = 5500$ s$^{-1}$ and $\Omega = 6100$ s$^{-1}$  in Fig. 3c.
It has been noted in Ref.[39] that  this local minimum in M-R curve at a fixed
angular velocity is "strictly" connected to the back bending phenomenon. We 
also find such a connection in this case and discuss it later. The curves
for $\Omega = 5500$ s$^{-1}$ and $\Omega = 6100$ s$^{-1}$  in Fig. 3c exhibit
two local maxima and they are denoted by solid circles. Here the parts of
fixed angular velocity curves where compact star masses decrease as a function 
of radius or central energy density between two local maxima, are not unstable.
This may be understood by studying the stability of fixed baryon number 
sequences. For example, it will be evident from the
stability analysis of the fixed baryon number curve HKQ, III which touches
the fixed angular velocity curve with $\Omega = 5500$ s$^{-1}$ 
around the local minimum.
We discuss this stability analysis in the following paragraph. 
It is noted here that the backward 
turning portions of curves III and IV in Fig. 3c are actually unstable parts. 

In Fig. 4, we have shown the stability (angular momentum versus central energy
density) of fixed baryon number sequences for HK and HKQ cases. The curve for 
HK case corresponds to the baryon number $N_b =2.09 \times 10^{57}$ whereas it 
is $N_b =2.45 \times 10^{57}$ for HKQ case. These fixed baryon number sequences 
for HK and HKQ cases are shown by curve III in Fig. 3b and 
in Fig. 3c respectively. In Fig. 4, we note that angular momentum, $J$, 
initially 
decreases with central energy density for both cases. This is shown by solid 
lines. It implies that the fixed baryon number sequences are stable in these 
regions as demanded by Eqn.(2). After the stable parts, there are unstable 
regions where angular momentum increases with central energy density. The 
unstable parts are represented by dotted lines. 
From the stability analysis of HKQ, III curve, we find that it is the stable 
part of the curve which touches the fixed angular velocity sequence with
$\Omega = 5500$ s$^{-1}$ around the local minimum in Fig. 3c. 
On the other hand, we find an interesting
structure in HK, III curve. After the unstable part in HK, III curve, there is a
stable region followed by an unstable region. This second stable part in HK 
case is the outcome of the high density behaviour of the EoS 
which actually gives rise to the third family of compact stars in Fig. 2 and 
Fig. 3b.

A rotating neutron star slowly loses its energy and angular momentum through
electromagnetic and gravitational radiation with time. However, it conserves 
the total
baryon number during this evolution. Therefore, fixed baryon number sequences
or evolutionary sequences provide valuable informations about isolated rotating
stars. In the following paragraphs, we analyse normal and supramassive sequences
of Figures 3.
We show the behaviour of angular velocity with angular momentum for CHM, HK
and HKQ cases in 
Fig. 5a, Fig. 5b and Fig. 5c respectively. The mass shedding limit sequence in 
each figure is shown by a light solid line. Also we have plotted normal and 
supramassive sequences in these figures. The stable parts of various sequences
are displayed by dark solid lines and the unstable parts by dotted lines. 
It is worth mentioning here that the stability of each curve is determined by Eqn. (2).
For CHM case, we have shown two normal sequences by curves I and II 
and a supramassive sequence by curve III. We find that neutron
stars along normal sequences spin down as they lose angular momentum. However,
the compact star along the supramassive sequence spins up with angular
momentum loss. This was already noted by various groups \cite{Cook,Ster95}. 
Similarly we plot normal and supramassive sequences 
for HK and HKQ cases in Fig. 5b and Fig. 5c. 
The most interesting result is obtained from curve II of Fig. 5b. In this case, we 
have an
unstable part followed by another stable part. The stable region beyond the
unstable part may be attributed to the high density behaviour of EoS for HK 
case. Already we have discussed that this EoS gives rise to
a stable third family branch of compact stars beyond the neutron star branch
in the static limit and fixed angular velocity sequences. In Fig. 5b,
we also note that the compact star on the second stable
branch rotates faster.

Now we display moment inertia (I) versus angular velocity ($\Omega$) for HK 
and HKQ cases in Fig. 6a and Fig. 6b respectively. In both figures, normal 
sequence (curve I) and supramassive sequences (curves III and IV) are 
displayed. In Fig. 6a and Fig. 6b, we 
find that the moment of inertia always diminishes with decreasing angular 
velocity for normal sequences. In Fig. 6a, 
supramassive sequences have both stable and unstable parts denoted by solid and
dotted lines respectively. For curve III of Fig. 6a, the moment of 
inertia initially decreases as the star rotates slowly along the stable part 
of the sequence. But the compact star spins up just before the beginning of 
the unstable part as it has been already observed for curve III of 
Fig. 5b. After the unstable region, there is a stable part followed by
an unstable part in curve III of Fig. 6a. The neutron star in the second stable
part spins
up with the loss of moment of inertia. On the other hand, 
the neutron star always spins up along the stable part of curve 
IV in Fig. 6a. 

In Fig. 6b, the supramassive sequence denoted by curve III exhibits an 
interesting structure. After the initial spin down of the neutron star along 
this sequence, there is a spin up followed by another spin down. This is known 
as back bending phenomenon. We observe that this back bending phenomenon is 
definitely connected to the appearance of the local minimum in the fixed 
angular velocity sequences in Fig. 3c. For example, the fixed baryon number 
curve III of Fig. 6b actually passes through the local minimum of the 
fixed angular velocity curve with $\Omega$ = 5500 $s^{-1}$ of Fig. 3c. Such a 
feature was earlier observed for an EoS involving a strong first order 
hadron-quark phase transition \cite{Web} and for the EoS including hyperons 
degrees of freedom \cite{Zdu}. In our calculation, the EoS for HKQ case
first undergoes the hadron-antikaon condensed matter phase transition and 
then the antikaon condensed-quark matter phase transition. It is the phase 
transition to quark matter which is responsible for the back bending in curve
III of Fig. 6b. Though we have a first order hadron-antikaon condensed matter 
phase transition in HK case, we do not find any such event in this case as
it is evident from the curves of Fig. 6a. It has
been stressed that J versus $\Omega$ curve provides more insight into the back 
bending phenomenon than I versus $\Omega$ curve \cite{Zdu}. For the former 
curve tells us whether the back bending segment contains stable configurations 
or not. For HKQ case, the back bending phenomenon occurs in the stable part of
the curve. 

\section{Summary and conclusions}

We have studied the effects of different forms of exotic matter such as 
hyperons, Bose-Einstein condensates of antikaons and quarks on rotating 
neutron stars. For this calculation, we adopted three equations of state. The
CHM case is described by a chiral model and this EoS does not involve 
any phase transition. On the other hand, the EoS for HK case involves a first
order phase transition from hadronic to $K^-$ condensed matter and then a 
second order $\bar K^0$ condensation whereas the EoS for HKQ case undergoes 
successive first order phase transitions from hadronic to $K^-$ condensed 
matter and then to quark matter. In both cases, the hadronic phase is described
by a relativistic field theoretical model. 

We have calculated various properties of rotating neutron stars with those 
equations of state. The maximum gravitational mass in static and mass shedding
limit sequences for HK case are smaller than that of HKQ case. For HK case, 
there exists a stable third family branch of compact stars beyond the
neutron star branch in the static sequence.

Next we have investigated fixed angular velocity sequences for the above 
mentioned equations of state. With increasing rotation, the fixed 
angular velocity curves became more flattened due to softening in EoS. For HK 
case, we 
find that the third family branch disappears for larger angular velocities. 
On the other hand, for HKQ case, the flattening in the fixed angular velocity 
sequences exhibit minima. We have seen that these minima are actually connected
to the back bending phenomenon. 

We have further computed fixed baryon number normal and supramassive sequences.
Using these evolutionary sequences, we have studied the stability and the 
behaviour of angular momentum and moment of inertia with angular velocity for
different equations of state. For HK case, we observe that the supramassive 
sequence has a second stable segment followed by another unstable
part. This second stable part is the result of the high density behaviour of 
the EoS for HK case which is responsible for the third family solution. 
It is found that neutron stars on the second stable branch rotate faster. 
For HKQ case, we observe the 
interesting back bending phenomenon in the moment of inertia versus angular
velocity curve. And the back bending segment contains stable configurations of
neutron stars. Phase transitions to antikaon condensed and quark matter in 
neutron star interior could be probed through the observation of back bending 
phenomenon.

\centerline{\bf Acknowledgements}

This work is supported by the Department of Science and Technology (DST), India
and German Academic Exchange Service (DAAD), Germany under a project based
Personnel Exchange Programme (sanction order no. INT/DAAD/P-94/2003). 

\newpage 

\begin{table}

\caption {Maximum gravitational masses ($M_G/M_{\odot}$), equatorial radii (R) 
and their 
corresponding central energy densities ($\epsilon_c$) for static ($P$ = 0) 
and Keplerian limit (P=$P_K$ = $2{\pi}/{\Omega_K}$) with different EoS, 
where $P_K$ is the Kepler period in milliseconds. The second row for HK case 
denotes those of the third family branch for the static case.
\label{tab1}}
\vskip 0.5 cm
\begin{tabular}{|ccccc|}
\hline
&&&&\\ 
{EoS}&${P (ms) }$&${\epsilon_c (10^{15} g/cm^3)}$&${M_G/M_{\odot}}$&${R (km)}$ \\ 

&&&&\\ \hline
{HK}&{0}&{1.34}&{1.569}&{13.28}\\

&{0}&{2.69}&{1.553}&{10.98}\\

{}&{1.0276}&{1.04}&{1.944}&{19.09}\\
&&&&\\

{HKQ}&{0}&{2.28}&{1.752}&{12.10}\\

{}&{0.8062}&{1.96}&{2.084}&{16.46}\\

&&&&\\
{CHM}&{0}&{1.99}&{1.636}&{11.72}\\

{}&{0.8289}&{1.66}&{1.973}&{16.52}\\

\end{tabular}
\end{table}
\newpage 

\vspace{2cm}

{\centerline{
\epsfxsize=12cm
\epsfysize=14cm
\epsffile{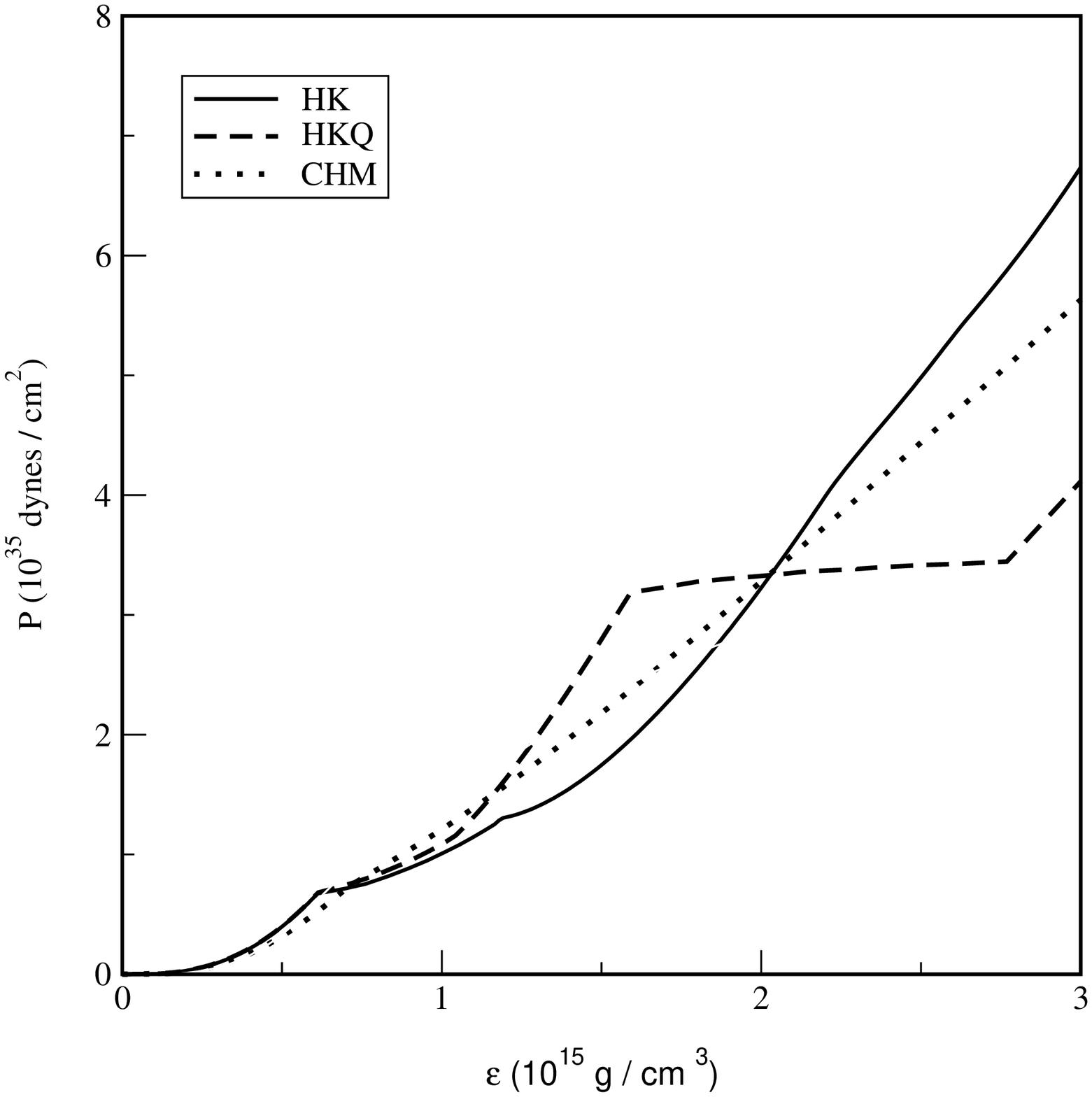}
}}

\vspace{1.0cm}

\noindent{\small{
FIG. 1. The equation of state, pressure $P$ vs. energy density $\varepsilon$ 
for CHM, HK and HKQ cases. The results are for  $n$, $p$, hyperon and lepton 
matter (dotted line), phase transitions from hadronic matter to $K^{-}$ and 
$\bar K^0$ condensed matter (solid line)
and phase transitions from nuclear matter to $K^{-}$  condensed matter and then
to quark matter (dashed line)
calculated with antikaon optical potential depth at normal nuclear matter 
density of $U_{\bar K} = -160$ MeV, bag constant B$^{1/4}$ = 200 MeV and
strange quark mass $m_s = 150$ MeV. }}

\vspace{2cm}

{\centerline{
\epsfxsize=12cm
\epsfysize=14cm
\epsffile{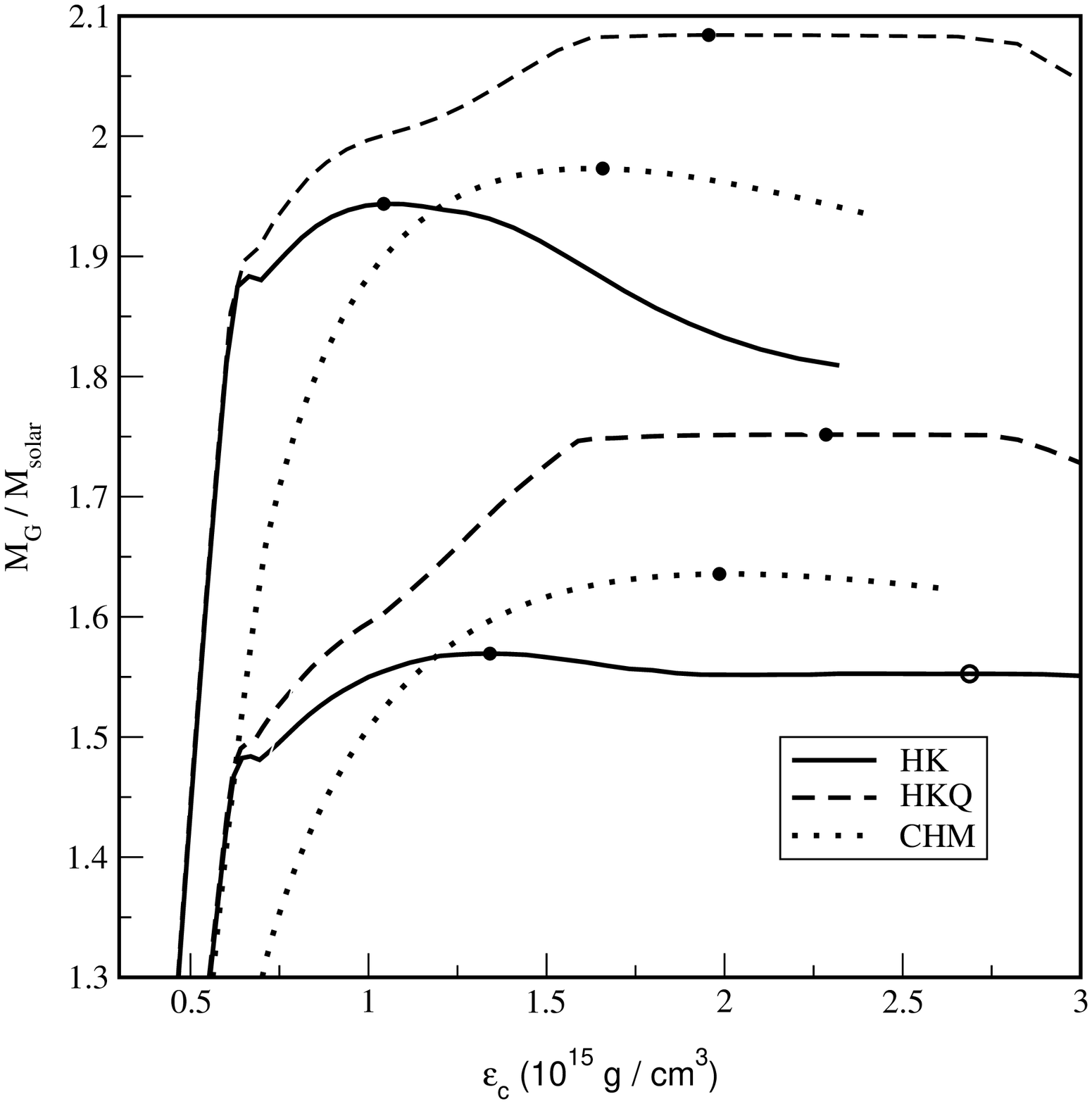}
}}

\vspace{1.0cm}

\noindent{\small{
FIG. 2. The gravitational mass for static compact star sequence and 
mass shedding limit sequence of 
rotating neutron stars are plotted with central energy 
density for equations of state shown in FIG. 1. The different lines have the 
same meaning as in FIG. 1.
\vspace{0.5cm}

\vspace{2cm}

{\centerline{
\epsfxsize=12cm
\epsfysize=14cm
\epsffile{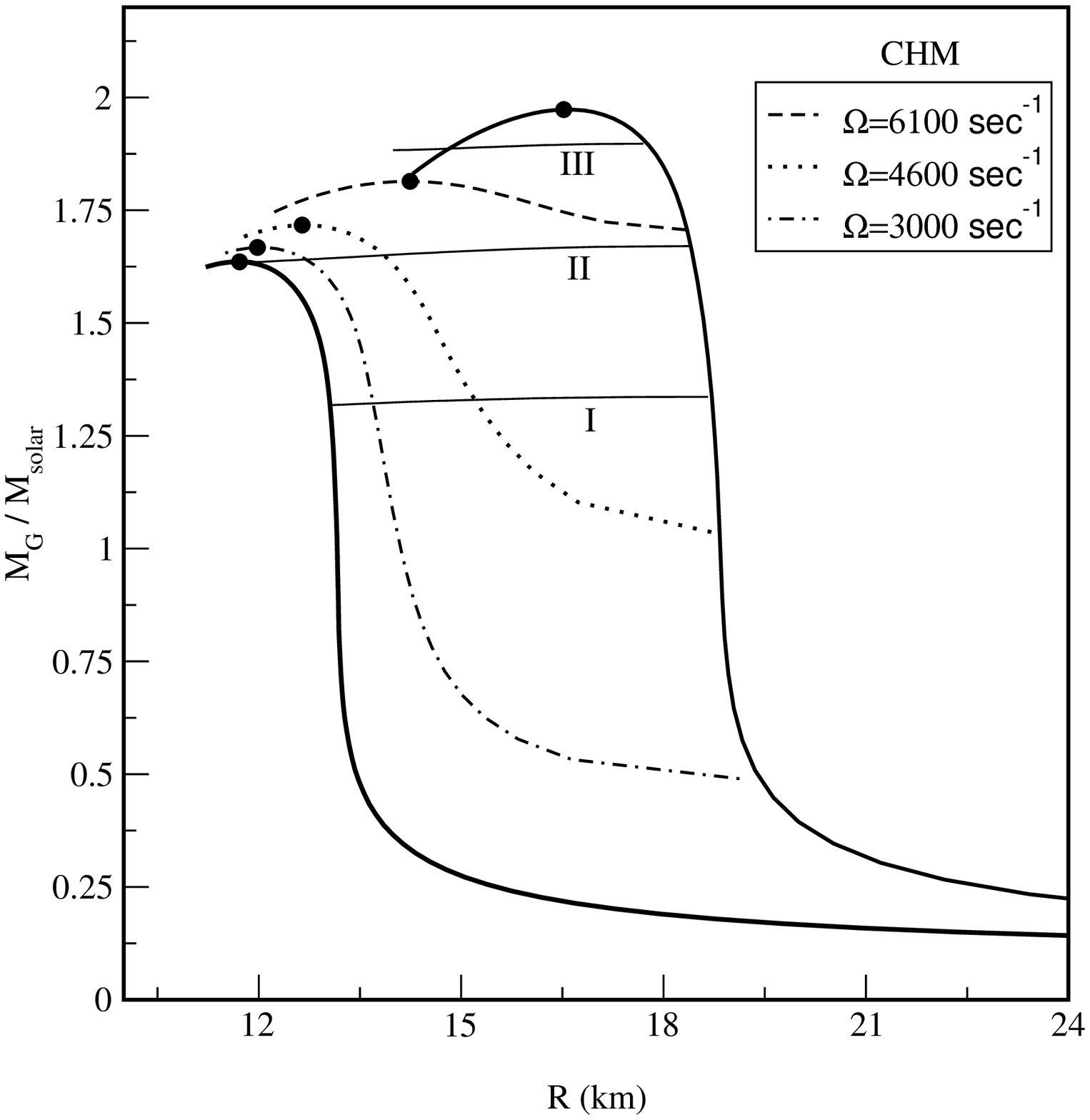}
}}

\vspace{1.0cm}

\noindent{\small{
FIG. 3a. The gravitational mass as a function of equatorial radius for CHM 
case. The extreme left and right bold curves show the static and mass shedding 
limit sequence respectively. Also, fixed angular velocity ($\Omega$) sequences 
are shown here. Horizontal lines are fixed baryon number sequences. Curves I 
and II are normal sequences and curve III is a supramassive sequence.
\vspace{0.5cm}

\vspace{2cm}

{\centerline{
\epsfxsize=12cm
\epsfysize=14cm
\epsffile{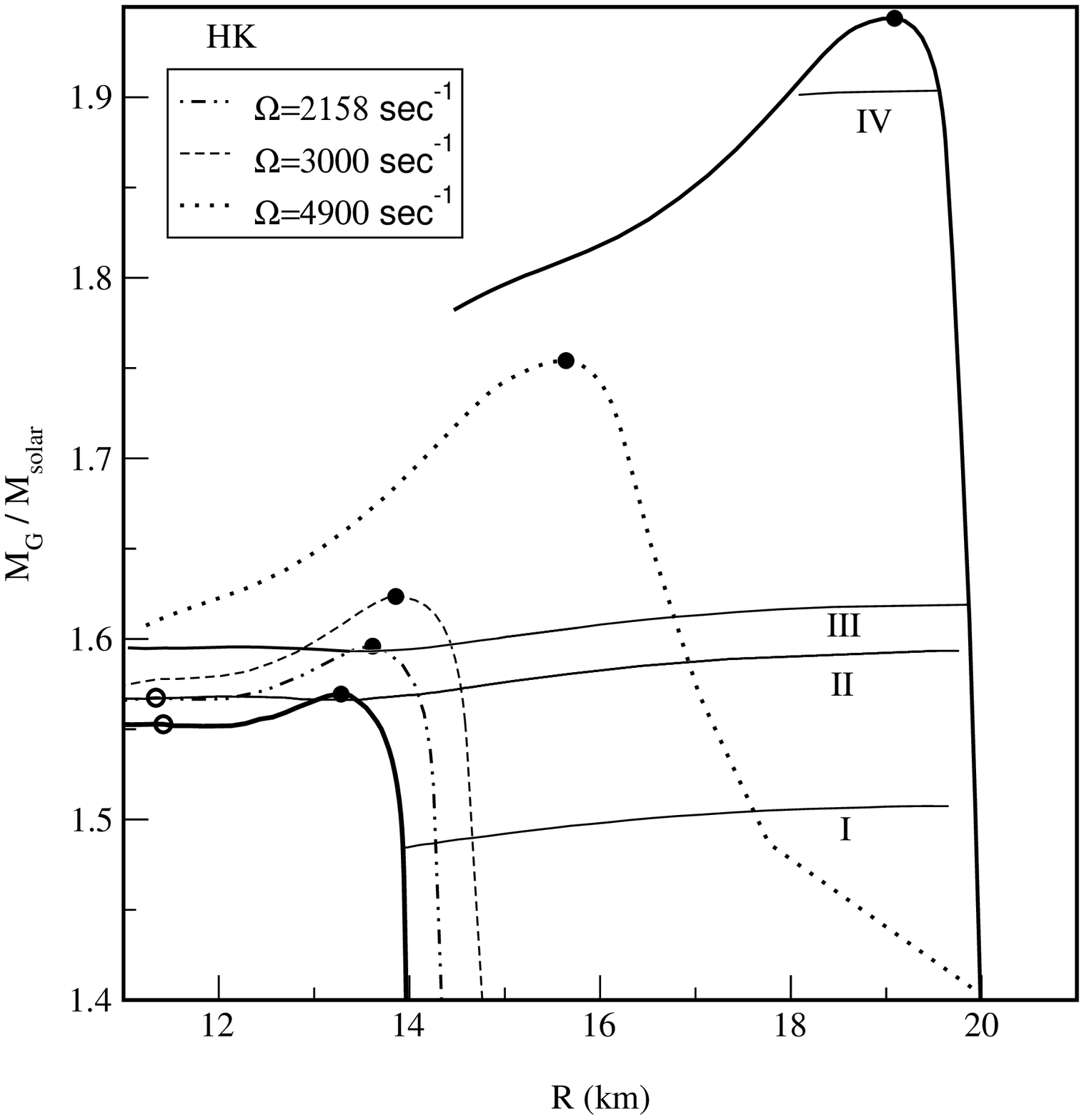}
}}

\vspace{1.0cm}

\noindent{\small{
FIG. 3b. Same as FIG. 3a but for HK case. Here curves I and
II are normal sequences and curves III and IV are supramassive sequences.
\vspace{0.5cm}

\vspace{2cm}

{\centerline{
\epsfxsize=12cm
\epsfysize=14cm
\epsffile{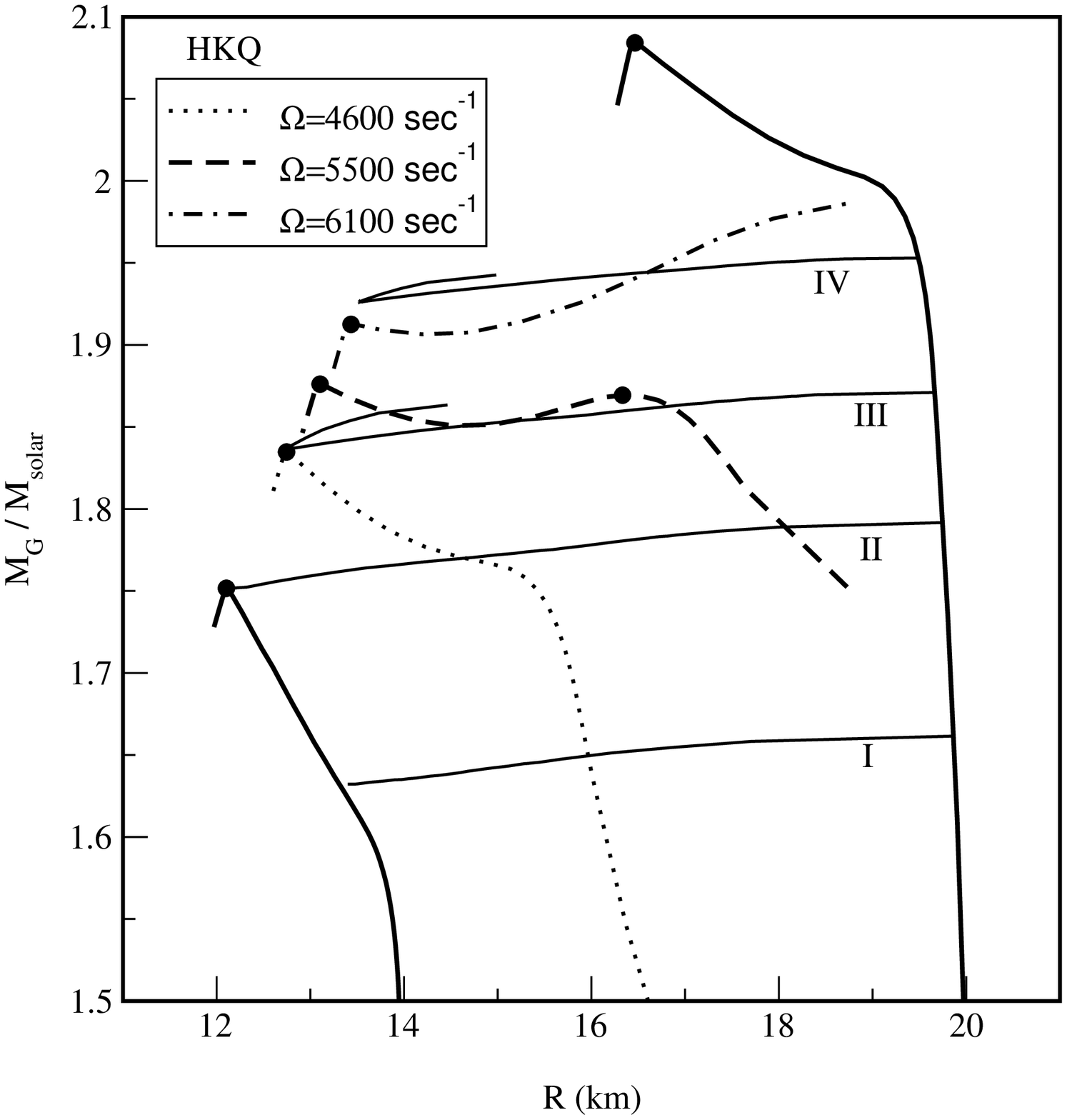}
}}

\vspace{1.0cm}

\noindent{\small{
FIG. 3c. Same as FIG. 3b but for HKQ case. 
\vspace{0.5cm}

\vspace{2cm}

{\centerline{
\epsfxsize=12cm
\epsfysize=14cm
\epsffile{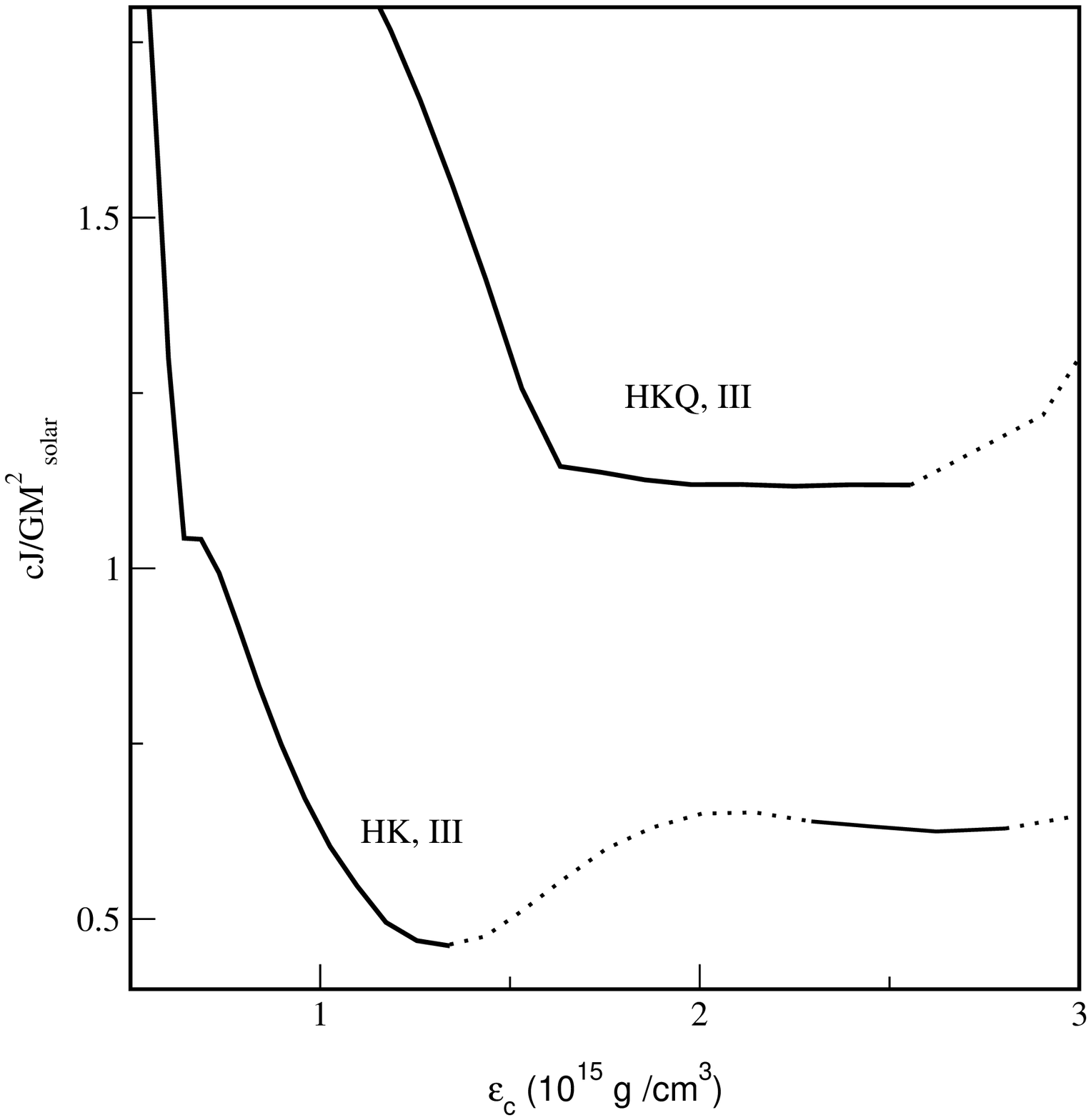}
}}

\vspace{1.0cm}

\noindent{\small{
FIG. 4. The angular momentum versus central energy density for fixed baryon 
number supramassive sequences of HK and HKQ cases is plotted. 
\vspace{0.5cm}

\vspace{2cm}

{\centerline{
\epsfxsize=12cm
\epsfysize=14cm
\epsffile{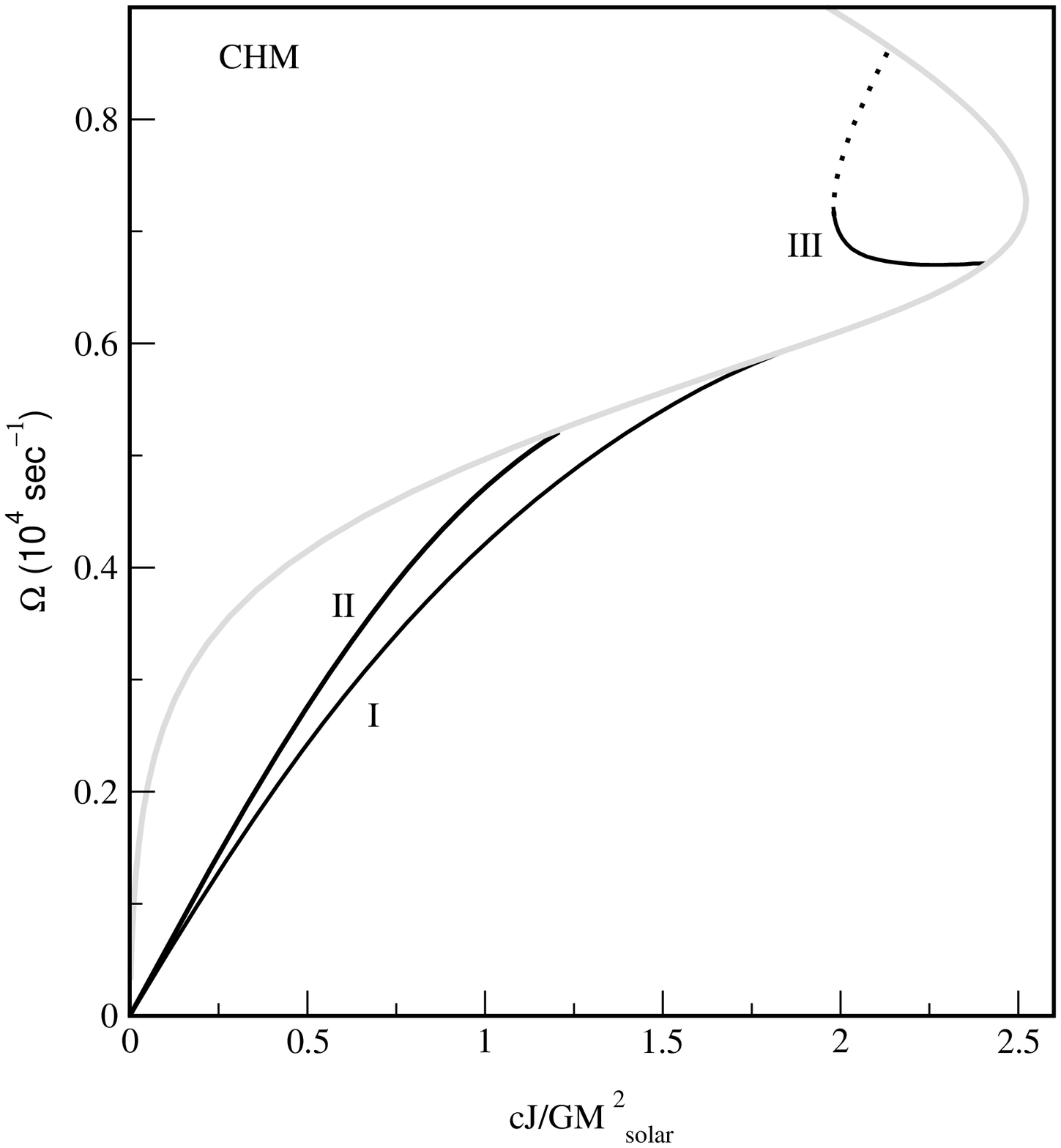}
}}

\vspace{1.0cm}

\noindent{\small{
FIG. 5a. The angular velocity versus angular momentum for CHM case. 
All labels are as indicated in Fig. 3a. The light solid line denotes the mass shedding limit 
sequence.
\vspace{0.5cm}

\vspace{2cm}

{\centerline{
\epsfxsize=12cm
\epsfysize=14cm
\epsffile{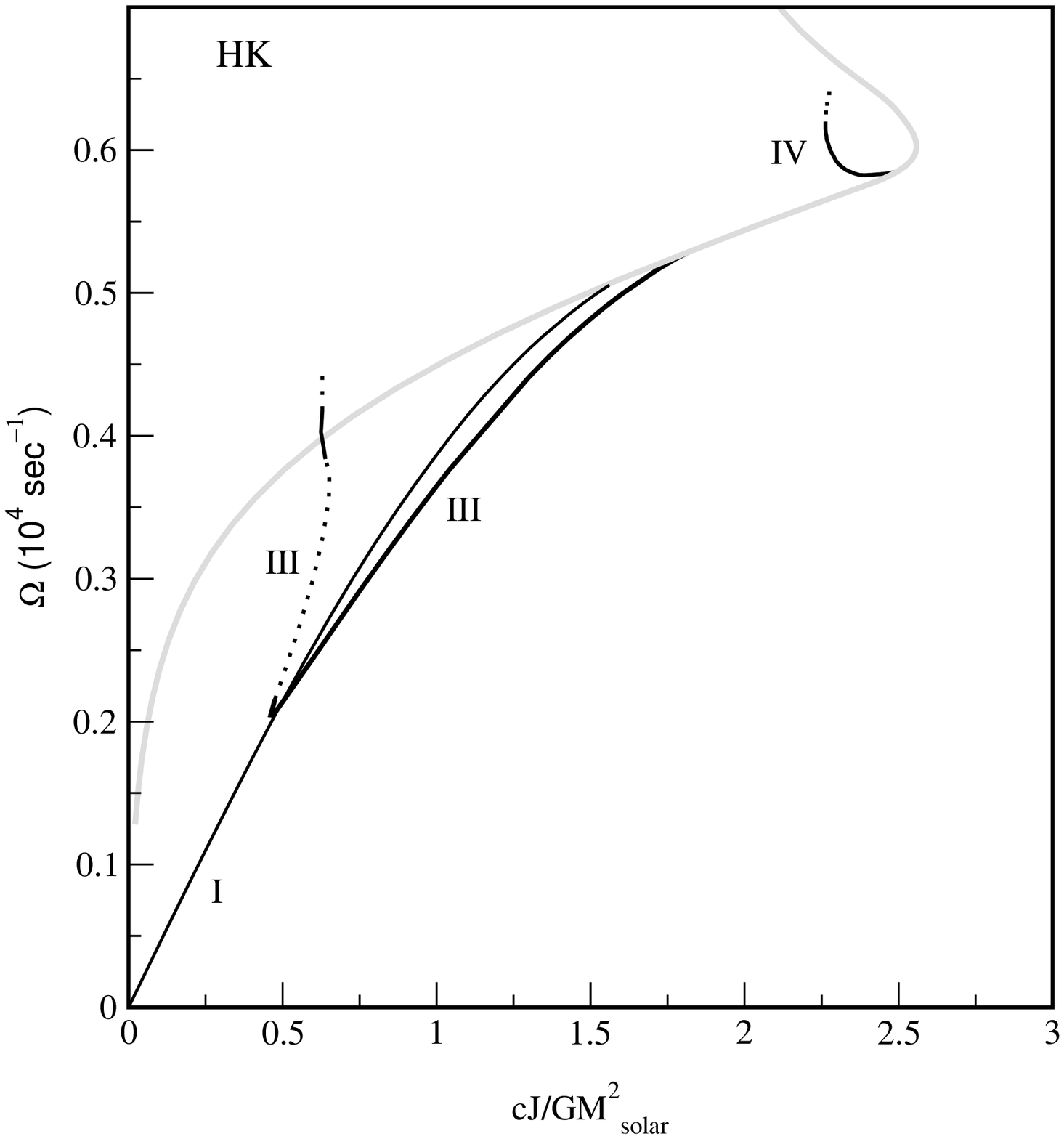}
}}

\vspace{1.0cm}

\noindent{\small{
FIG. 5b. Same as FIG. 5a but for HK case. All labels are as indicated in Fig. 3b. 
\vspace{0.5cm}

\vspace{2cm}

{\centerline{
\epsfxsize=12cm
\epsfysize=14cm
\epsffile{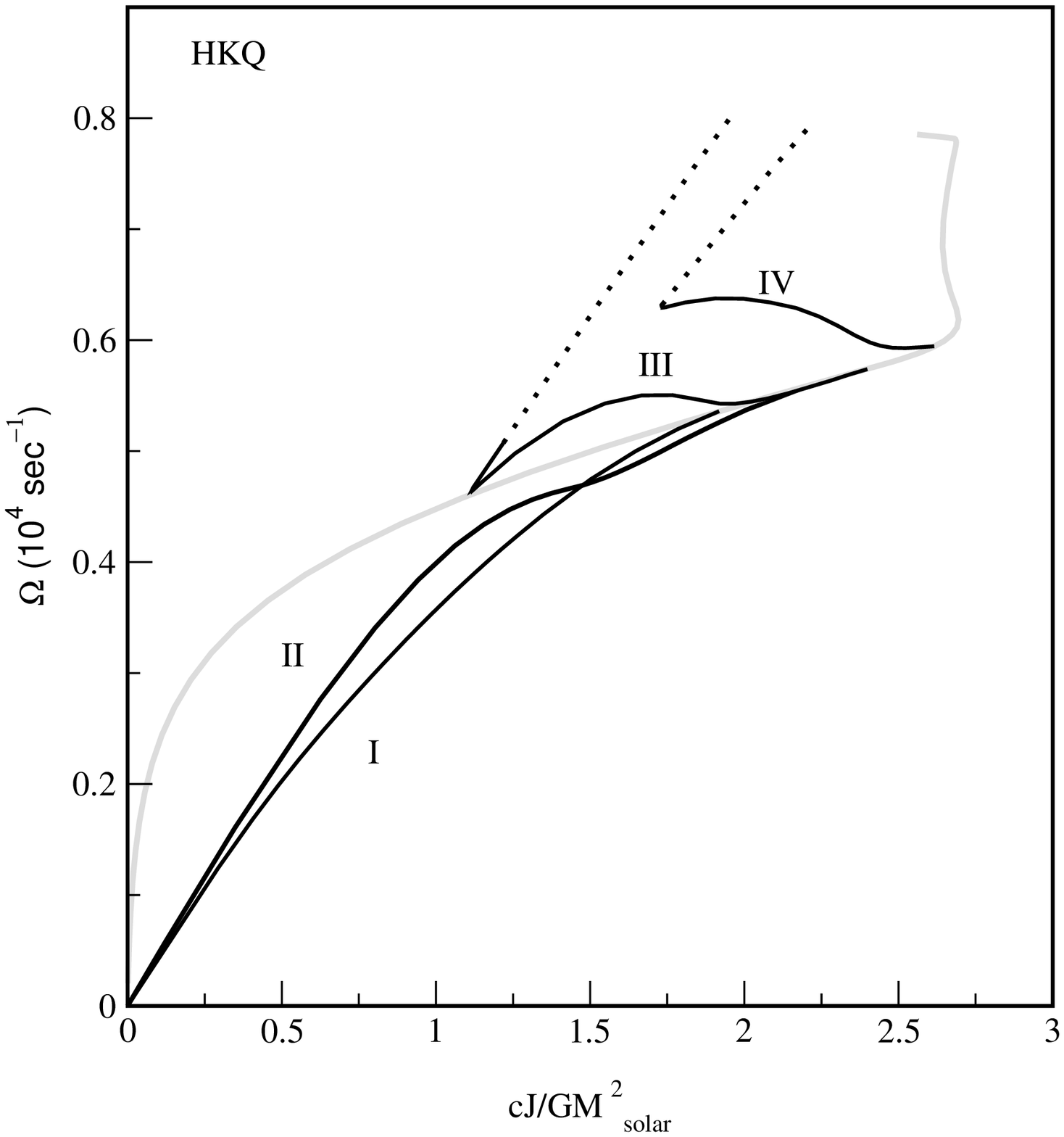}
}}

\vspace{1.0cm}

\noindent{\small{
FIG. 5c. Same as FIG. 5a but for HKQ case. All labels are as indicated in Fig. 3c. 
\vspace{0.5cm}

\vspace{2cm}

{\centerline{
\epsfxsize=12cm
\epsfysize=14cm
\epsffile{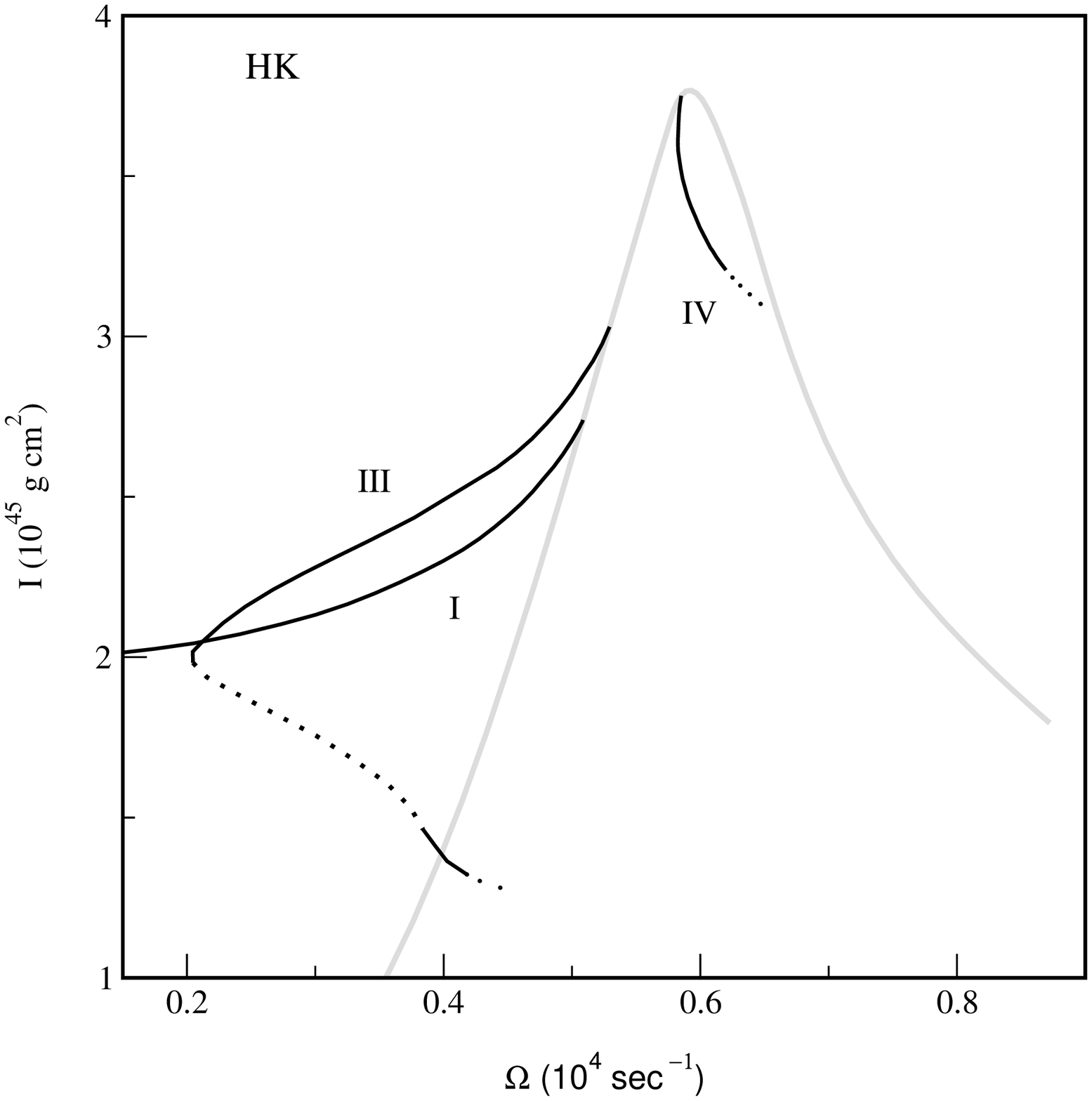}
}}

\vspace{1.0cm}

\noindent{\small{
FIG. 6a. The moment of inertia versus angular velocity for HK case. 
All labels are as indicated in Fig. 3b.
The light solid line implies the mass shedding limit.
\vspace{0.5cm}

\vspace{2cm}

{\centerline{
\epsfxsize=12cm
\epsfysize=14cm
\epsffile{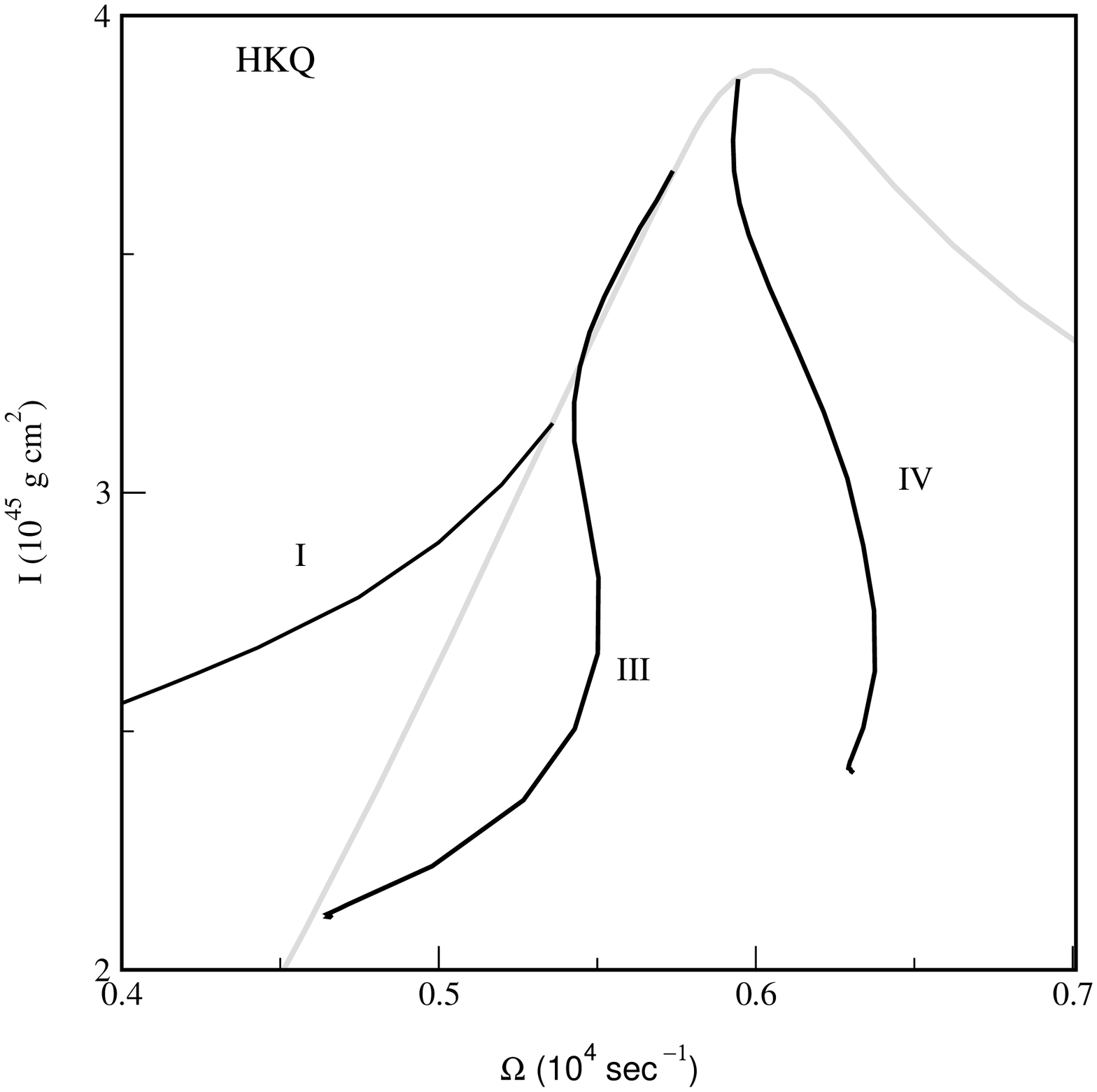}
}}

\vspace{1.0cm}

\noindent{\small{
FIG. 6b. Same as FIG. 6a but for HKQ case. All labels are as indicated in Fig. 3c. 
\end{document}